\documentclass[runningheads, envcountsame, a4paper]{llncs}
\usepackage{algorithm,algorithmic}
\usepackage[misc]{ifsym}
\usepackage{graphicx}
\usepackage{hyperref}
\usepackage{amsmath}
\usepackage{xcolor}
\usepackage{subcaption}
\usepackage{microtype}

\begin{document}
\title{Topic Aware Contextualized Embeddings for High Quality Phrase Extraction\thanks{This research work is supported by Extramarks Education India Pvt. Ltd. (an education technology company), SERB, FICCI (PM fellowship for doctoral research) and TiH Anubhuti (IIITD).}}
%
%
%
\author{Venktesh V \Letter \and
Mukesh Mohania \and
Vikram Goyal}
\tocauthor{Venktesh ~ V,  Mukesh  ~ Mohania,  Vikram ~ Goyal}
\toctitle{Topic Aware Contextualized Embeddings for High Quality Phrase Extraction}
\authorrunning{Venktesh V. et al.}
%
\institute{Indraprastha Institute of Information Technology, Delhi \\
\email{\{venkteshv, mukesh, vikram\}@iiitd.ac.in}
}
%
\maketitle              
\begin{abstract}
Keyphrase extraction from a given document is the task of automatically extracting salient phrases that best describe the document. This paper proposes a novel unsupervised graph-based ranking method to extract high-quality phrases from a given document. We obtain the contextualized embeddings from pre-trained language models enriched with topic vectors from Latent Dirichlet Allocation (LDA) to represent the candidate phrases and the document. We introduce a scoring mechanism for the phrases using the information obtained from contextualized embeddings and the topic vectors. The salient phrases are extracted using a ranking algorithm on an undirected graph constructed for the given document. In the undirected graph, the nodes represent the phrases, and the edges between the phrases represent the semantic relatedness between them, weighted by a score obtained from the scoring mechanism. To demonstrate the efficacy of our proposed method, we perform several experiments on open source datasets in the science domain and observe that our novel method outperforms existing unsupervised embedding based keyphrase extraction methods. For instance, on the SemEval2017 dataset, our method advances the F1 score from 0.2195 (EmbedRank) to 0.2819 at the top 10 extracted keyphrases. Several variants of the proposed algorithm are investigated to determine their effect on the quality of keyphrases. We further demonstrate the ability of our proposed method to collect additional high-quality keyphrases that are not present in the document from external knowledge bases like Wikipedia for enriching the document with newly discovered keyphrases. We evaluate this step on a collection of annotated documents. The F1-score at the top 10 expanded keyphrases is 0.60, indicating that our algorithm can also be used for ’concept’ expansion using external knowledge.

\end{abstract}
\section{Introduction}
Keyphrases are the salient terms in a document that serve as summaries of the document. They play an important role in many text processing applications like document clustering, classification, information retrieval \cite{kim-etal-2013-applying,phrasier} and text generation. Automatic Keyphrase Extraction (AKE) task is a crucial component of these applications to obviate the need for manual extraction of keyphrases.  
In addition to the above-mentioned applications, we are primarily interested in the applications of AKE in online learning platforms.
Learning contents in such learning platforms are tagged at the topic level for accessibility. However, a topic can be further divided into concepts that enable linking of related learning content and easy navigation through the learning contents.
In this paper, concepts are characterized as keyphrases as they describe the content of a document. We posit that the Automatic Keyphrase Extraction (AKE) from learning contents can help to index the massive collection of learning contents in online learning platforms enabling better accessibility of the learning contents.
 Automatic Keyphrase Extraction is a well studied problem \cite{hasan-ng-2014-automatic}.  Unlike the supervised methods, the unsupervised methods do not require annotated documents and rely on in-corpus statistical information for extracting keyphrases.
 In most of the unsupervised keyphrase extraction methods  \cite{bougouin-etal-2013-topicrank,liu-etal-2010-automatic,liu-etal-2009-clustering,mihalcea-tarau-2004-textrank} the candidate phrases are represented by a word graph formed based on a co-occurrence window and then ranked. Recent unsupervised methods like EmbedRank \cite{bennani-smires-etal-2018-simple} leverage representation learning methods that help to capture the semantic relatedness between the phrases and the document. 
 
We propose a novel unsupervised method to automatically extract keyphrases from a given document. In this method, the candidate phrases and the given document are represented in a continuous vector space by combining the contextual embeddings and the topical information from LDA \cite{10.5555/944919.944937} to strengthen the associations between phrases that occur in similar contexts and also represent similar topics. Then a graph based ranking algorithm where the nodes are represented by phrases than words is employed to rank the phrases.
The proposed unsupervised method helps to capture two important characteristics needed for keyphrases: \textit{coherence} and \textit{informativeness}. We posit that the selected phrases are \textit{coherent} if they convey a consistent idea \cite{seisa} and they are \textit{informative} if they convey the core ideas discussed in the document. In the proposed method, \textit{coherence} is captured as the cosine similarity computed between the embeddings of the candidate phrases. The \textit{informativeness} of a phrase is captured as the cosine similarity computed between the embeddings of the candidate phrase and the document. The proposed algorithm outperforms existing unsupervised AKE methods. For instance, on SemEval2017  dataset  our  method  advances  the  F1  score  from  0.2195  (EmbedRank)  to \textbf{0.2819}.

Following are the core technical contributions of our paper:
 \vspace{1mm}
\begin{itemize}
\item We propose a new topic aware representation method to represent the phrases and the document. To the best of our knowledge, this representation method has not been applied to the task of keyphrase extraction.
\item We propose a graph based ranking method with a new scoring mechanism that captures the  \textit{informativeness} and \textit{coherence} measures by using the proposed representation method.
\item We apply our algorithm on the task of enriching the set of extracted keyphrases with new keyphrases which are not present in the source document by using external knowledge sources like Wikipedia.
\end{itemize}
The code and data can be found at \url{https://github.com/VenkteshV/Unsupervised_keyphrase_extraction_CoTagRank_ECIR_2022}.

 \section{Related Work}

In this section, we discuss the existing unsupervised AKE methods \cite{papagiannopoulou2019review} and also the current advancements in the vector representation methods.
\subsection{Unsupervised keyphrase extraction}
Graph based ranking methods are the most popular among the unsupervised AKE algorithms. 
Graph based AKE methods were introduced in the seminal work TextRank \cite{mihalcea-tarau-2004-textrank}. It constructs a uniformly weighted graph for the given text where an edge connects word types only if they co-occur within a window of specified size. The SingleRank \cite{10.5555/1620163.1620205} algorithm was an extension to the TextRank algorithm where the edges were assigned a weight equal to the number of times word types co-occur. The WordAttractionRank \cite{wang2014corpus} algorithm is similar to SingleRank with one difference. It incorporates the distance between the word embeddings into the weighting scheme for the edges between the words. As keyphrases usually appear at the beginning of the document, the PositionRank \cite{florescu2017positionrank} algorithm and the MultiPartiteRank \cite{boudin-2018-unsupervised} algorithm assigns weights to nodes (words and phrases respectively) favouring the terms appearing at initial positions in the text. One of the shortcomings of these approaches (except MultiPartiteRank) is that they rank the phrases by aggregating the scores of the constituent words. This can lead to uninformative candidate phrases being ranked higher just because one of the constituent words has a higher score.

Several existing approaches like TopicRank \cite{bougouin-etal-2013-topicrank} have leveraged topical information for ranking phrases. Another algorithm that leverages the topical information is the TopicalPagerank (TPR) \cite{liu-etal-2010-automatic} algorithm. The TPR method runs TextRank for each topic where the topics are obtained using LDA \cite{10.5555/944919.944937}. Another extension to TPR is the Salience Rank algorithm \cite{teneva-cheng-2017-salience} which introduces word salience metric to balance between topic and corpus specificity of words. 

In contrast to the graph based methods, the EmbedRank \cite{bennani-smires-etal-2018-simple} algorithm is an embedding based AKE method that represents both documents and candidate phrases as vectors using document embedding methods like Sent2Vec \cite{Pagliardini2018UnsupervisedLO}. The vector representations help to rank candidates by computing cosine similarity between the phrase vectors and the document vector.

\subsection{Contextualized vector representations}
Distributed representations that capture the semantic relationships \cite{10.5555/2999792.2999959} have helped to advance many NLP tasks. But all the classical embedding methods like fasttext \cite{bojanowski-etal-2017-enriching}, GloVe \cite{pennington-etal-2014-glove}  generate fixed vectors for even polysemous words irrespective of the context in which they occur..

The Bidirectional Encoder Representation from Transformers (BERT) \cite{DBLP:journals/corr/abs-1810-04805} is one of the current state of the art methods that uses a mechanism called \textit{attention} \cite{10.5555/3295222.3295349}. The attention mechanism helps to encode a word using other positions in the input sequence that would lead to a better representation for the word. The Sentence BERT \cite{reimers-gurevych-2019-sentence} model was proposed to generate useful sentence embeddings by fine-tuning BERT. Another transformer based sentence encoding model is the Universal Sentence Encoder (USE) \cite{cer-etal-2018-universal}  that has been specifically trained on semantic textual similarity task.
In our experiments, we demonstrate that our novel representation using topic distribution based vectors from LDA (Latent Dirichlet Allocation) and USE embeddings performs better than USE or BERT embeddings in isolation. 

\section{Methodology}

In this section, we describe the proposed extraction algorithm, CoTagRank. First, the candidate phrases are extracted based on the Part Of Speech (POS) tags using the pattern {${<NN.*|JJ>*<NN.*>}$} \cite{10.5555/1620163.1620205}. Then the phrases and the document are projected to a continuous vector space and phrases are ranked as discussed in the following sections.
\subsection{Vector representations for the phrases and the document}

The primary goal of our algorithm is to extract the candidate phrases that best describe the document. We define two measures for achieving this goal namely \textit{coherence} and \textit{informativeness}. The coherence measure can be seen as an indicator that the candidate phrases represent a consistent idea. The informativeness measure can be seen as an indicator as to whether the phrases convey the core ideas discussed in the document. We posit that the above two measures can be captured by leveraging the topical information. Hence, we give a novel vector representation mechanism that combines topical information with the embeddings obtained from the state of the art contextualized embedding methods. We leverage contextualized embeddings to handle polysemous words. An example of polysemy can be seen in the following two sentences : "Consider an \textbf{imaginary} box", "An \textbf{imaginary number} is a complex number". In the above two sentences, the word "imaginary" has different meanings. Contextualized embeddings capture the context of usage of the word and hence produce different vector representations for the same word depending on the context.

The phrase representations are obtained by combining the contextualized embeddings of the phrases with the topic vectors of their constituent words obtained from LDA. The LDA is a generative probabilistic model in which each word in document d is assumed to be generated by sampling a topic from \textit{d}'s topic distribution $\theta^d$ and then sampling a word from the distribution over words denoted by $\phi^t$ of a topic. We use pre-trained Universal Sentence Encoder (USE) to obtain contextualized embeddings for both the phrases and the sentences as it has been pre-trained on the Semantic Text Similarity (STS) task.  This representation method helps in bringing the phrases that are semantically related and having similar topic distributions closer in the vector space. This implies that the phrases that are both semantically related and represent similar topics would have a higher coherence measure (cosine similarity between phrase representations). The phrase representations are obtained in the following manner:

         \[   LE(CP) = \sum_{w \in CP}[p(w |t_1 ), p(w|t_2)...]\]
                  \vspace{-1em}

          \begin{equation}        CPE = concat(LE(CP), CE(CP))      \end{equation}   
where LE represents LDA embeddings, CP represents a candidate phrase, CE represents contextualized embeddings, and CPE represents candidate phrase embeddings. The vector $[p(w |t_1 ), p(w|t_2)...]$ represents the word-topic probabilities that are derived from the word distributions $\phi^t$ over the topics. 
Similarly the document representation can be obtained by combining topic distribution of the document with the contextualized embeddings of the document sentences. The document representation is obtained as follows:
         \[   LE(doc) = [p(t_1 |d ), p(t_2 |d )...]\]
         \vspace{-1em}
\begin{equation}         
DE = concat(LE(doc), CE(doc))      
\end{equation}
where, the vector $[p(t_1 |d ), p(t_2 |d )...]$ represents the document-topic probabilities and $DE$ represents the document embedding.
Latent Dirichlet Allocation (LDA) is run only once on the corpus of documents and not for every document.
The vector representations obtained as described are leveraged in the graph based ranking step to compute final scores for the candidate phrases.

\subsection{Graph based ranking }
In this subsection, we explain our graph based candidate phrases ranking method in detail.


It differs from traditional methods like TextRank, SingleRank and PositionRank. We construct an undirected graph with the candidate phrases as the vertices instead of words. Constructing the graph in this manner circumvents the overgeneration errors that occur when the phrases are ranked by aggregating the scores of the top ranked words. Hence using word graph based AKE methods may result in an uninformative phrase being assigned a high score just because one of the constituent words has a higher score. The edges connecting the phrases are weighted by the semantic relatedness (cosine similarity) computed between the vector representations of the phrases. The vector representations for the phrases are obtained as described in the previous subsection. The edges are formed between the phrases (nodes) that co-occur in the original text within a specified window size (tunable parameter). We demonstrate that when the window size is set to the maximum value for forming a complete graph, we get the maximum performance. The completeness nature of the graph has the benefit of connecting phrases together that may not co-occur together but having similar topic distributions. As mentioned in the previous subsection, our goal is to rank those phrases higher that are coherent and informative for the document. The \textit{coherence} measure is represented by the edge weights of the graph. The \textit{informativeness} measure for each phrase ($P_a$)  is the normalized cosine similarity between the document and the phrase representations computed as follows:
\begin{equation}  
                     \begin{split}
                          n\_Sim(P_a,doc) = \frac{Sim(P_a,doc) - min_{Pb \in P} (Sim(P_b, doc))  } { max_{P_b \in P} (Sim(P_b, doc))}           
                     \end{split}
                     \end{equation}

            where $n\_Sim$ is normalized cosine similarity, $Sim$ is the cosine similarity function and $doc$ represents the document. Then the similarity metric is obtained as : 
\begin{equation}
 \begin{split} 
                     F\_Sim(P_a,doc) = \frac{n\_Sim(P_a,doc) - \mu(n\_Sim(P, doc))} {\sigma(n\_Sim(P, doc))}     
                     \end{split}
  \end{equation}

where $P$ is the set of phrases. The $F\_Sim$ function returns the final cosine similarity metric obtained after normalization and standardization of cosine similarities. The function $n\_Sim$ given a set of embeddings of the phrases and a document embedding as inputs returns a vector of normalized cosine similarities. Each element in the output vector is the normalized similarity between the corresponding embedding in the set and the document embedding. In Equation (3), $n\_Sim(Pa, doc)$, $P_{a}$ denotes a set of embeddings having only one element. Whereas in Equation (4), $n\_Sim(P, doc)$, P has multiple embeddings.

The  goal is to find the phrases that maximize the objective:
      \[ Obj = \lambda S_{coh}(P) + (1- \lambda) S_{inf}(P,doc)\]
where, $S_{inf}$ denotes the function that returns the \textit{informativeness} measure computed using Equation 4.

The $S_{coh}$ is the function that computes the \textit{coherence} measure. The $S_{coh}$ denotes a function that takes a set of embeddings of phrases (P) and outputs a vector of cosine similarities computed between embeddings of all possible pairs of distinct phrases in the set. The parameter $\lambda$ balances the importance given for $S_{coh}$ and $S_{inf}$ factors.

Iteratively optimizing the above objective is similar to random walk based approaches. Hence maximizing the above objective can be done as follows:


Every candidate phrase in the graph is ranked by:
                \begin{equation}  \begin{split}  R(p_i) = \lambda  \sum_{j:p_j->p_i} \frac{e(p_i,p_j)}{OutDeg(p_j)} R(p_j) + (1 - \lambda)  {S_{inf}(p_i)} \end{split} \end{equation}

where $e(p_i,p_j)$ denotes the weight of the edges between the phrases ($p_i$ and $p_j$) (\textit{coherence}) and $S_{inf}(p_i)$ is the \textit{informativeness} score that helps in biasing the random jump to phrases (vertices) that are closer to the document in the vector space.

We explore several variants of \textit{CoTagRank} in \textbf{Section 4}.

 \begin{table*}
\caption{Statistics of the datasets used}

\label{tab1}

\centering
\begin{tabular}{c|p{1.2cm}|p{1.4cm}|p{2.3cm}|p{2cm}|p{2.7cm}}
\small
Dataset & Domain & \# of docs & \# of tokens/doc & \# of gold keys& \# of gold keys/doc \\ \hline \hline

\bf Inspec & Science & 2000 & 128.20 &29230&14.62
\\ 
\hline
\bf SemEval2017  & Science  &  493& 178.22 & 8969 &18.19\\\hline
\bf SemEval2010 & Science & 243 &  8332.34 &4002& 16.47 \\ \hline
\end{tabular}
\vspace{-4mm}
\end{table*}
\section{Experiments and Results}
This section, discusses the experimental setup and results. 

\subsection{Datasets}
We evaluate our algorithm on standard datasets like Inspec \cite{hulth-2003-improved}, SemEval 2017 \cite{augenstein-etal-2017-semeval} and SemEval 2010 \cite{kim-etal-2010-semeval} for keyphrase extraction. We choose SemEval2017 and Inspec as they contain documents of short length resembling the learning content in e-learning platforms. We also show the performance of our method on a dataset containing long documents such as SemEval2010. The statistics of the datasets are shown in Table 1. Since our algorithm is completely unsupervised, we evaluate on all the documents in each of these datasets.
\begin{table} [h!]
\small
\centering
 \caption{Performance comparison. \textsuperscript{\textdagger} indicates significance at 0.01 level (t-test).\\ \textsuperscript{$\ddagger$} indicates that effect size $> 0.2$. }
 \fontsize{7.5}{7}\selectfont

\centering
\begin{tabular}{p{2cm}|p{6cm}|p{1.3cm}|p{1.2cm}|p{1.2cm}}

\hline \bf Dataset  & \bf Method & \bf P@10 & \bf R@10 & \bf F1@10 \\  \hline\hline
SemEval2017 & TopicalPageRank &  0.3523 & 0.2098  & 0.2543 \\
            & MultiPartiteRank & 0.2972 & 0.1758 & 0.2133 \\
            & SingleRank &  0.3428 & 0.2040  & 0.2474 \\
            & TextRank &  0.1848 & 0.1069  & 0.1326 \\
            & WordAttractionRank & 0.2566 & 0.1482 & 0.1815\\ \cline{2-5}
            & EmbedRank &  0.3061 & 0.1801  & 0.2195 \\
            & EmbedRankSentenceBERT &  0.3329 & 0.1982  & 0.2404 \\
            & EmbedRankSentenceUSE &   0.3286 & 0.1965  & 0.2381 \\\cline{2-5}
            & $CoTagRank$ (our algorithm) &  \bf 0.3911$\dagger$$\ddagger$ & \bf 0.2324$\dagger$$\ddagger$  & \bf 0.2819$\dagger$$\ddagger$ \\
            & $CoTagRankSentenceUSE$ (our algorithm) & 0.3860 & 0.2290 & 0.2779\\
            & CoTagRanks2v (our algorithm) & 0.3379 & 0.1990 & 0.2424\\
            & $CoTagRankWindow$ (w=10) (our algorithm) & 0.3797 & 0.2253 & 0.2734\\
            & $CoTagRankWindow_{positional}$ (our algorithm) &0.3793 & 0.2250& 0.2731 \\\hline
Inspec      & TopicalPageRank &  0.2724 & 0.2056  & 0.2260 \\
            & MultiPartiteRank & 0.2210 & 0.1710  & 0.1865\\
            & SingleRank &  0.2694 & 0.2044  & 0.2239 \\
            & TextRank &  0.1408 & 0.1020  & 0.1234 \\
            & WordAttractionRank & 0.1778 & 0.1437 & 0.1516\\\cline{2-5}
            & EmbedRank &   0.2732 &  0.2034  &  0.2259 \\
            & EmbedRankSentenceBERT &  0.2663 & 0.1970  & 0.2188 \\
            & EmbedRankSentenceUSE &   0.2748 & 0.2049  & 0.2267 \\\cline{2-5}
            & $CoTagRank$ (our algorithm) &   \bf 0.2984$\dagger$$\ddagger$ & \bf 0.2213$\dagger$$\ddagger$  &  \bf 0.2454$\dagger$$\ddagger$   \\
            & $CoTagRankSentenceUSE$ (our algorithm) & 0.2881 & 0.2150 & 0.2377\\
            & CoTagRanks2v (our algorithm) & 0.2372 & 0.1807 & 0.1983\\
            & $CoTagRankWindow$ (w=10) (our algorithm) & 0.2747 & 0.2062 & 0.2275\\ 
            & $CoTagRankWindow_{positional}$ (our algorithm) &0.2750 & 0.2062& 0.2276 \\ \hline
SemEval2010 & TopicalPageRank &  0.0477 & 0.0293  & 0.0359 \\
            & MultiPartiteRank &  \bf 0.1757$\dagger$$\ddagger$ &  \bf 0.1118$\dagger$$\ddagger$  &  \bf 0.1352$\dagger$$\ddagger$\\
            & SingleRank &   0.0457 & 0.0277  &  0.0341 \\
            & TextRank &  0.0321 & 0.0199  & 0.0243 \\
            & WordAttractionRank & 0.0835 & 0.0531 & 0.0641\\\cline{2-5}
            & EmbedRank &   0.0128 &  0.0082  &  0.0099 \\
            & EmbedRankSentenceBERT &   0.0230 & 0.0137  & 0.0170 \\
            & EmbedRankSentenceUSE &   0.0379 & 0.0241  & 0.0292 \\\cline{2-5}
            & $CoTagRank$ (our algorithm) &   0.0695 &  0.0434  &  0.0530 \\
            & $CoTagRankSentenceUSE$ (our algorithm) & 0.0671 & 0.0418 & 0.0511\\
            & CoTagRanks2v (our algorithm) & 0.0267 & 0.0169 & 0.0204\\
            & $CoTagRankWindow$ (w=10) (our algorithm) & 0.1337 & 0.0867 & 0.1042\\
            & $CoTagRankWindow_{positional}$ (our algorithm) & \textcolor{darkgray} {0.1494} & \textcolor{darkgray} {0.0970}& \textcolor{darkgray} {0.1165} \\ \hline
SemEval2010  & TopicalPageRank &  0.1745 & 0.1100  & 0.1336 \\
(abstract and & MultiPartiteRank &  0.1646 &  0.1044  &  0.1263\\
 intro)            & SingleRank &   0.1580 & 0.0998  &  0.1211 \\
            & TextRank &  0.1140 & 0.0719  & 0.0872 \\
            & WordAttractionRank & 0.1481 & 0.0949 & 0.1145\\\cline{2-5}
            & EmbedRank &   0.0654 &  0.0407  &  0.0496 \\
            & EmbedRankSentenceBERT &   0.0844 & 0.0521  & 0.0638 \\
            & EmbedRankSentenceUSE &   0.1243 & 0.0760  & 0.0933 \\\cline{2-5}
            & $CoTagRank$ (our algorithm) & 0.1811 & 0.1134 & 0.1380\\
            & $CoTagRankSentenceUSE$ (our algorithm) & 0.1786 & 0.1121 & 0.1363\\
            & CoTagRanks2v (our algorithm) & 0.0852 & 0.0518 & 0.0636\\
            & $CoTagRankWindow$ (w=10) (our algorithm) & 0.1856 & 0.1170 & 0.1419\\
            &$CoTagRankWindow_{positional}$ (our algorithm) & \bf 0.1909$\dagger$$\ddagger$ & \bf 0.1203$\dagger$$\ddagger$ & \bf 0.1459$\dagger$$\ddagger$ \\
\hline

\end{tabular}

\vspace{-2mm}
\end{table}
\subsection{Baselines and Variants of the Proposed Method}
In this section, we describe the variants of the proposed CoTagRank algorithm and other baselines. In the proposed CoTagRank algorithm, a complete graph is formed from the phrases. The phrases and the document are represented by combining the contextualized embeddings from Universal Sentence Encoder (USE) (512-dimensional) \cite{cer-etal-2018-universal} and topical vectors from Latent Dirichlet Allocation (LDA)\footnote{We leveraged the sklearn implementation for LDA \url{https://scikit-learn.org/}}. The number of topics K was set at 500 when running LDA. \\
We compare $CoTagRank$ with several variants such as:
\begin{itemize}
    \item $CoTagRankWindow$: This algorithm is a variant of \textit{$CoTagRank$} where only the phrases that co-occur in the text within a window of the specified size are connected in the graph. While $CoTagRank$ forms a complete graph of phrases, CoTagRankWindow provides a tunable parameter, the window size $w$, which determines the edges formed between phrases. The vector representation and the ranking method is the same as explained in Section 3.2.
    \item CoTagRanks2v: This algorithm is similar to $CoTagRank$ with respect to complete graph formation and ranking using Equation 5. However, in \textit{CoTagRanks2v} the static sentence representation method like Sent2Vec \cite{Pagliardini2018UnsupervisedLO} is used to project the phrases and the document to a continuous vector space. 
    \item $CoTagRankSentenceUSE$: A variant of the $CoTagRank$ where the document and phrase are encoded using only Universal Sentence Encoder yielding 512-dimensional representations.

\end{itemize}

We also consider two variants of EmbedRank such as EmbedRankSentenceBERT and EmbedRankSentenceUSE where \textit{bert-base-nli-stsb-mean-tokens}
from sentence-transformers\footnote{\url{https://huggingface.co/sentence-transformers/bert-base-nli-stsb-mean-tokens}} and Universal Sentence Encoder are used respectively as vector representation methods. We compare the performance of the proposed algorithms with strong baselines such as EmbedRank (Sent2Vec)\footnote{\url{https://github.com/swisscom/ai-research-keyphrase-extraction}}, SingleRank and other unsupervised AKE methods\footnote{\url{https://bit.ly/369Ycg7}}. 
\subsection{Results and Discussion}
The performance comparison of the algorithms are as shown in Table 2. The measures used to evaluate the algorithms are Precision, Recall and F1-score. The metrics were computed using trec-eval\footnote{\url{https://github.com/usnistgov/trec_eval}}. Since the original implementation of EmbedRank did not provide an evaluation script, we use trec-eval to compute the metrics for EmbedRank and we observe different results from those reported in the original EmbedRank paper\footnote{Our results are close to the implementation in the project \url{https://bit.ly/2IbbyjT} which also uses trec-eval and the original EmbedRank implementation}.

As shown in Table 2, the CoTagRank outperforms existing graph based and embedding based unsupervised methods on two of the three datasets and on the third dataset, we get comparable results to the MultiPartiteRank algorithm. The performance gain obtained using CoTagRank over EmbedRankSentenceBERT, EmbedRankSentenceUSE and $CoTagRankSentenceUSE$ demonstrates the advantage of fusing topical information with the contextualized embeddings rather than leveraging just contextualized embeddings for phrase and document representations.

\begin{figure}[h!]
\centering
\begin{subfigure}{0.32\linewidth}
\includegraphics[width=1.2\linewidth]{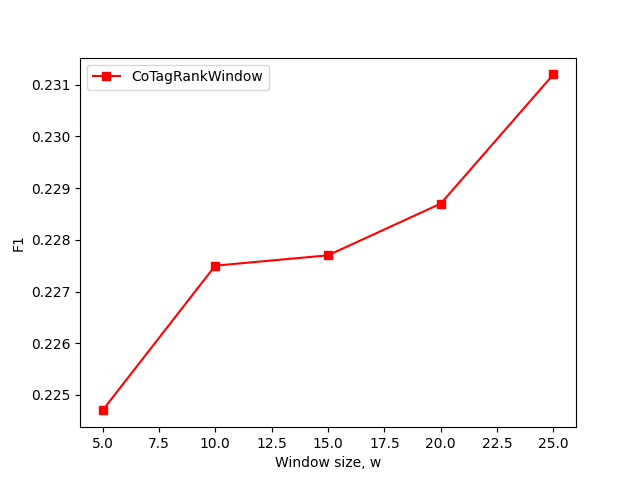}
\caption{{ F1-scores for varying window size on Inspec}}
\end{subfigure}
\hspace{0.3em}
\begin{subfigure}{0.32\linewidth}
\includegraphics[width=1.2\linewidth]{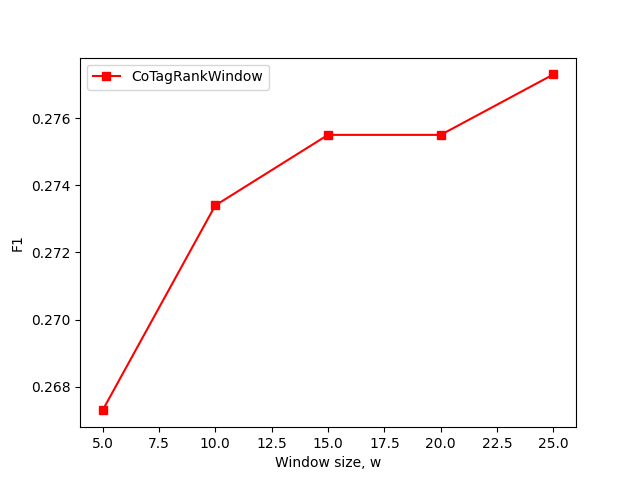}
\caption{  F1-scores for varying window size on SemEval2017  }
\end{subfigure}%
\hspace{0.3em}
\begin{subfigure}{0.31\linewidth}
\includegraphics[width=1.2\linewidth]{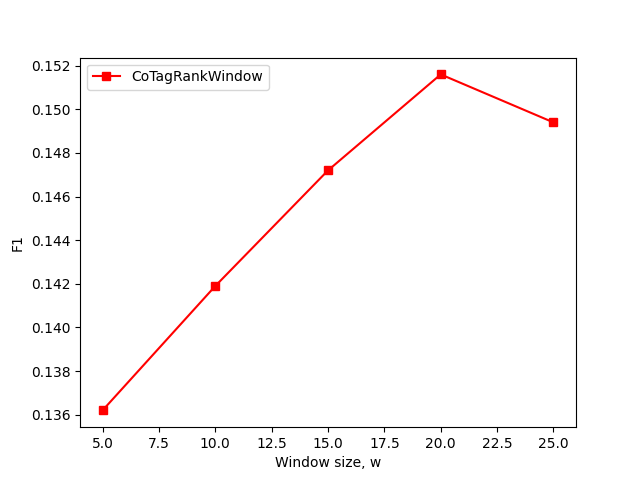}
\caption{  F1-scores with varying window sizes on SemEval2010 (abstract and intro)  }
\end{subfigure}%
\hspace{0.4em}
\label{subfig}
\begin{subfigure}{0.31\linewidth}
\includegraphics[width=1.2\linewidth]{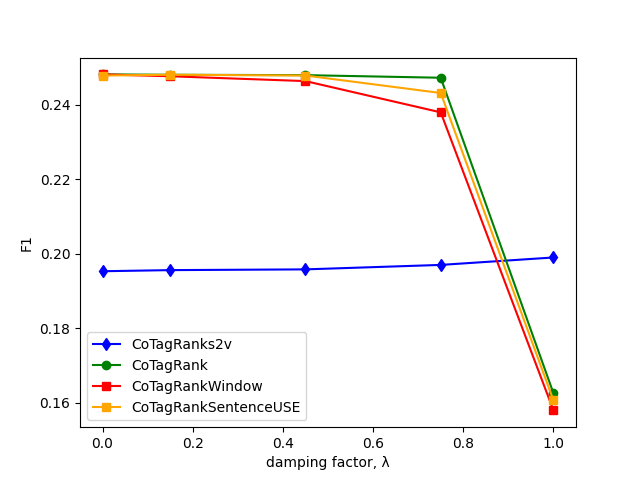}
\caption{ F1-scores for different values of $\lambda$ on Inspec}
\end{subfigure}
\hspace{0.6em}
\begin{subfigure}{0.31\linewidth}
\includegraphics[width=1.2\linewidth]{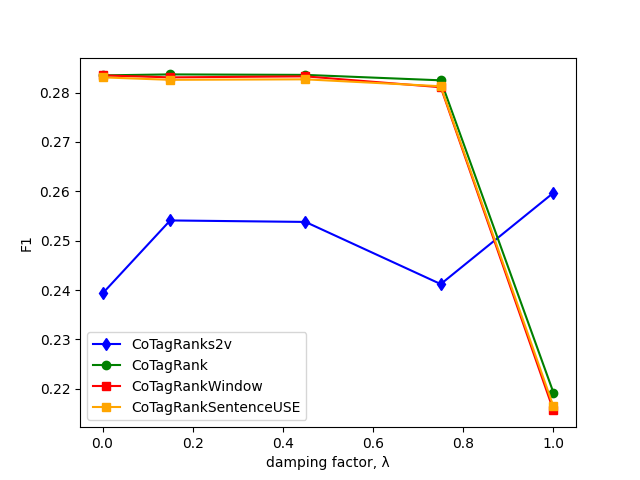}
\caption{ F1-scores for different values of  $\lambda$ on SemEval2017 }
\end{subfigure}%
\hspace{1em}
\begin{subfigure}{0.31\linewidth}
\includegraphics[width=1.2\linewidth]{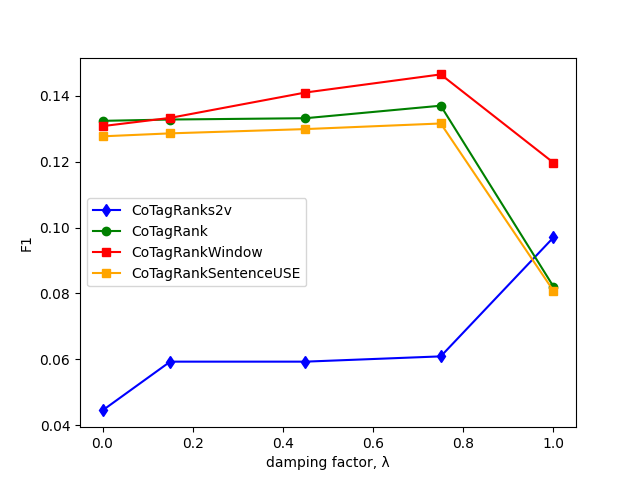}
\caption{ F1-scores for varying $\lambda$ on SemEval2010 (abstract and intro) }
\end{subfigure}%

\caption{ Performance comparison for different hyperparameters}
\label{fig}
\vspace{-2mm}
\end{figure}

However, on long documents, the MultiPartiteRank outperforms all other methods. The MultiPartiteRank algorithm leverages the position of the candidate phrases in the document as a feature that leads to the gain in F1-score on long documents. This result is similar to the result reported in the EmbedRank paper \cite{bennani-smires-etal-2018-simple}. To overcome this limitation, the authors of EmbedRank propose a variant $EmbedRank_{positional}$, which includes the position of the candidate phrases as a feature to increase the performance on long documents. In contrast to the $EmbedRank_{positional}$ method, we were able to achieve a gain in performance by just tuning the window size in the CoTagRank algorithm. The results in Table 2 show that the performance of a variant of our proposed algorithm, $CoTagRankWindow$ with a window size of 10, is close to the performance of MultiPartiteRank. We also verify the intuition of positional bias by proposing $CoTagRankWindow_{positional}$ where we multiply the node weights by the inverse of the start position of the phrase in the input document. We observe that the performance with the positional bias is close to the  MultiPartiteRank algorithm. Additionally, we leverage the common knowledge that most keyphrases are located at the beginning of the document and hence perform keyphrase extraction only on "Abstract" and "Introduction" sections of every document in SemEval 2010, as shown in Table 2, which advances the F1 scores of CoTagRank and $CoTagRank_{positional}$ to \textbf{0.1380} and \textbf{0.1459} respectively surpassing MultiPartiteRank confirming the positional bias intuition. On average, this reduced version of the SemEval 2010 dataset still contains 550 tokens when compared to the number of tokens in Inspec and SemEval 2017, as observed in Table 1. 

However, since online learning contents like questions and video transcripts are usually short text documents, the expected performance of $CoTagRank$ on such documents is closer to the results observed for Inspec and SemEval 2017. We also perform statistical significance tests and observe that our results are significant at \textbf{($p<0.01$)} with effect sizes of \textbf{0.37}, \textbf{0.30} and \textbf{0.83} for F1 scores on SemEval2017, Inspec and SemEval2010 (abstract and intro) respectively.

\subsection{Effects of different hyperparameters}
In this section, we discuss the effect of varying hyperparameters such as window size (w), damping factor ($\lambda$) and number of topics in LDA (LDA embeddings dimension) of the $CoTagRank$ algorithm and its variants.
 We vary the window size hyperparameter $w$ with values 5, 10, 15, 20 and 25. The graphs in the Figure 1a and 1b show that the F1-score increases with the increase in window size for the $CoTagRankWindow$ algorithm. The window size can be set to the maximum value encompassing all phrases in the document forming a complete graph. This validates our claim that running the biased PageRank algorithm on a complete graph of phrases helps in producing high quality phrases. However, the same assumption may not hold good for longer documents, as evident from Figure 1c.
   \begin{figure*}[h!]
\begin{subfigure}{0.6\linewidth}
\includegraphics[width=1\linewidth,height=0.38\linewidth]{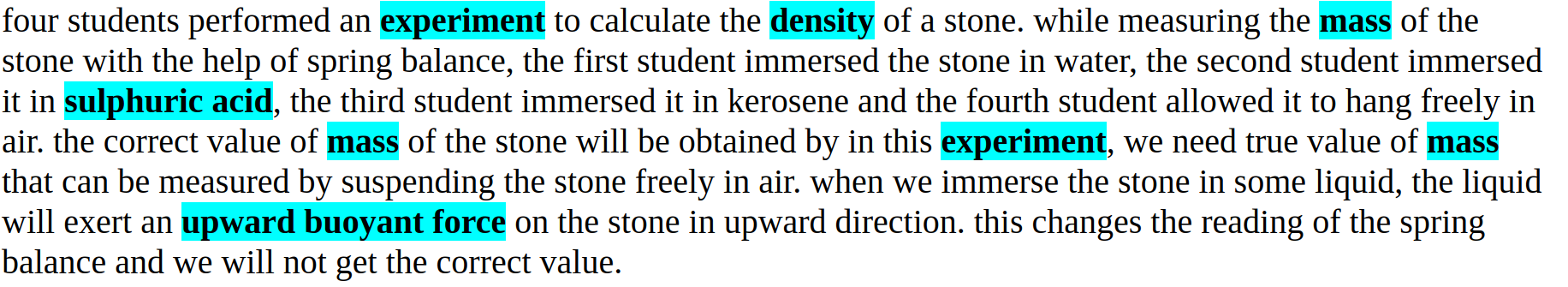}
\end{subfigure}
\begin{subfigure}{0.48\linewidth}
\includegraphics[width=1\linewidth]{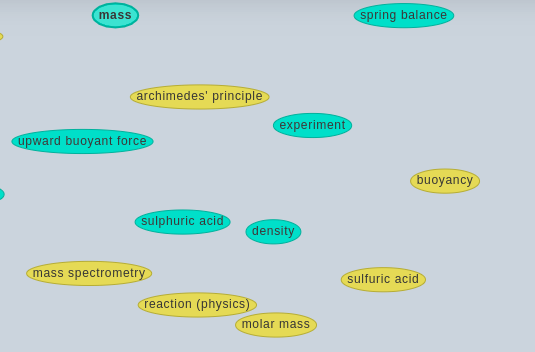}
\end{subfigure}

\caption{Keyphrase expansion results for an academic learning content}
\label{expansion}
\vspace{-2mm}
\end{figure*}
We observe that the performance of $CoTagRankWindow$ on SemEval 2010 (abstract and intro) increases with increase in window size drops a little at $w$ = 25. This indicates that forming a complete graph may not lead to the highest performance on longer documents.
We also vary the damping factor $\lambda$ in Equation 5. The values we experiment with are {0, 0.15, 0.45, 0.75, 1.0}. The graphs in the Figure 1d and 1e show that in  the proposed CoTagRank algorithm and in the variant $CoTagRankWindow$ the performance declines with an increase in the damping factor. When damping factor is set to 1, the $S_{inf}$ component in Equation 5 that contributes to \textit{informativeness} of the phrase becomes zero resulting in a drop in F1-score. The decrease in F1-score observed in the plots as the damping factor increases supports our claims of \textit{informativeness} and \textit{coherence} measures. However, we do not observe this trend in CoTagRanks2v. This may be due to the representation method used for the phrases, which do not contribute to the informativeness measure defined in this paper. From Figure 1f, we can observe that on SemEval2010 (abstract and intro) dataset, when $\lambda$ is set to 1, there is a drop in F1-score. However, when compared to the previous two graphs, we observe that the relative drop in F1-score is low. This maybe due to the length of the documents in this dataset when compared to short length documents in Inspec and SemEval 2017.

We also vary the number of topics (LDA embeddings dimension) and observe that $CoTagRank$ and \textit{CoTagRankWindow} achieves the highest performance when number of topics is set to 500. This is similar to the observation made in the TopicalPageRank paper \cite{liu-etal-2010-automatic} where the authors show that setting the number of topics to 500 gives the highest performance on the Inspec dataset.

\subsection{Keyphrase expansion results}
We further apply our algorithm to the task of keyphrase expansion to enrich the document with new keyphrases with the help of external knowledge sources like Wikipedia. This would help in linking related learning content in online platforms. 

The keyphrases extracted from the source document using the CoTagRank algorithm serve as seed set for the keyphrase expansion task.

We use the wrapper over MediaWiki API\footnote{https://pypi.org/project/wikipedia/} to extract relevant Wikipedia article titles for each keyphrase in the seed set. Then the expanded phrases are ranked using Equation 5. 
 \begin{table*} [hbt!] 
 \vspace{-4mm}
 \fontsize{9.8}{10.9}\selectfont
 \caption{\label{font-table} Performance comparison at top-10 expanded keyphrases }
 \small
\centering
\begin{tabular}{l|l|r|r|r}

\hline \bf Dataset  & \bf Method & \bf Precision & \bf Recall & \bf F1 \\  \hline\hline
Lecture transcripts & CoTagRank (our algorithm) &  \bf 0.5448 & \bf 0.7270  & \bf 0.6096\\
   (Khan academy)    & CoTagRanks2v & 0.2483 & 0.3424 & 0.2956 \\
                     & $CoTagRankSentenceUSE$ & 0.5207 & 0.6950 & 0.5949 \\ \hline

\end{tabular}
\vspace{-4mm}
\end{table*}

To demonstrate the effectiveness of this algorithm, we applied it for keyphrase expansion on 30 lecture transcripts collected from Khan academy in the science domain. The extracted phrases were given to two annotators who were undergraduate students in the Computer Science department familiar with the concepts. The task was to annotate the phrases as relevant to the document (1) or not relevant to the document (0).

The degree of agreement on relevance of keyphrases between the two annotators was measured using Cohen's kappa $\kappa$. We obtained a $\kappa$ of 0.535 denoting \textit{moderate} agreement between the annotators. A phrase is considered as a ground truth label only if both the annotators consider it to be relevant. We compute the Precision, Recall and F1 metrics as shown in Table 3. The F1 score of \textbf{0.6096} indicates that the proposed algorithm was able to retrieve relevant keyphrases from external knowledge sources. We observe that CoTagRanks2v and $CoTagRankSentenceUSE$ do not perform well in this task, indicating that the combination of contextualized embeddings and topic representations help in extracting better keyphrases from external knowledge sources.

Figure \ref{expansion} shows the results of running the proposed algorithm on an academic content from Khan academy. We observe that our algorithm was able to discover interesting phrases like \textit{Archimedes principle} though it was not present in the source document. The new keyphrases can help in linking related learning content, where the given question in Figure \ref{expansion} can be linked with a video explaining \textit{Archimedes principle}. We observed that none of the other algorithms were able to retrieve \textit{Archimedes principle}. This further reinforces the idea that apart from semantic relatedness between phrases that occur in similar contexts, their topic relatedness is also captured through our representation mechanism. The evaluation of the proposed algorithm on this corpus demonstrates that our algorithm could also enrich the existing set of keyphrases with new keyphrases using external knowledge sources like Wikipedia. 
\vspace{-2mm}

\section{Conclusions}
\vspace{-2mm}
In this paper, we proposed a novel representation and graph based ranking algorithm, CoTagRank, for keyphrase extraction. The algorithm is currently deployed to extract academic concepts from learning content in an online learning platform. We showed that our method outperforms existing state-of-the-art unsupervised keyphrase extraction methods in shorter texts and comparable performance on longer texts. In addition, forming a complete graph of phrases outperforms window based graph formation methods on short documents. We also demonstrated that including a simple positional bias helps further advance the performance of the algorithm on longer documents. In the future, we aim to incorporate positional embeddings and verify the performance on long texts.

\bibliographystyle{splncs04}
\bibliography{ecir}




\end{document}